\documentstyle[aas2pp4]{article}

\received{21 March 1997}

\lefthead{M.~M\'endez et al.}
\righthead{Kilohertz QPO and Atoll Source States in 4U~0614+09}

\begin{document}

\title{Kilohertz QPO and Atoll Source States in 4U~0614+09}

\author{M.~M\'endez\altaffilmark{1,2},
        M.~van der Klis\altaffilmark{1},
        J.~van Paradijs\altaffilmark{1,3},
        W.H.G.~Lewin\altaffilmark{4},
	F.K.~Lamb\altaffilmark{5},
	B.A.~Vaughan\altaffilmark{6},
	E. Kuulkers\altaffilmark{7},
	D. Psaltis\altaffilmark{5}}

\altaffiltext{1}{Astronomical Institute ``Anton Pannekoek'',
       University of Amsterdam and Center for High-Energy Astrophysics,
       Kruislaan 403, NL-1098 SJ Amsterdam, the Netherlands}

\altaffiltext{2}{Facultad de Ciencias Astron\'omicas y Geof\'{\i}sicas, 
       Universidad Nacional de La Plata, Paseo del Bosque S/N, 
       1900 La Plata, Argentina}

\altaffiltext{3}{Physics Department, University of Alabama in Huntsville,
       Huntsville, AL 35899, USA}

\altaffiltext{4}{Massachusetts Institute of Technology, Center for Space
       Research, Room 37-627, Cambridge, MA 02139, USA}

\altaffiltext{5}{Departments of Physics and Astronomy, University of
       Illinois at Urbana-Champaign, Urbana, IL 61801, USA}

\altaffiltext{6}{Space Radiation Laboratory, California Institute of
       Technology, MC 220-47, Pasadena CA 91125, USA}

\altaffiltext{7}{Astrophysics, University of Oxford, Nuclear and
       Astrophysics Laboratory, Keble Road, Oxford OX1 3RH, United
       Kingdom}

\begin{abstract}

We report three RXTE/PCA observations of the low-mass X-ray binary
4U~0614+09.  They show strong ($\sim$30\% rms) band-limited noise with
a cut-off frequency varying between 0.7 and 15~Hz in correlation with
the X-ray flux, $f_{\rm x}$.  We observe two non-simultaneous 11--15\%
(rms) kHz peaks near 728 and 629~Hz in the power spectra of two of our
observations when $f_{\rm x} \sim$10$^{-9}$erg~cm$^{-2}$~s$^{-1}$
(2--10~keV), but find no QPO ($<$6\% rms) when $f_{\rm x}$ is half
that.  We suggest that count rate may not be a good measure for $\dot
M$ even in sources as intrinsically weak as 4U~0614+09, and that QPO
frequency and noise cutoff frequency track $\dot M$ more closely than
count rate.  The QPO increases in rms amplitude from (11$\pm$1.3)\%
at 3 to (37$\pm$12)\% at 23~keV; the fractional amplitude of the
band-limited noise is energy-independent.  This suggests different
sites of origin for these two phenomena.  The spectrum of the
oscillating flux roughly corresponds to a black body with temperature
(1.56$\pm$0.2)~keV and radius (500$\pm$200)~m (other models fit as
well), which might indicate the oscillations originate at a small
region on the neutron star surface.

\end{abstract}


\section{Introduction}

Recently, kilohertz quasi-periodic oscillations (kHz QPOs) have been
discovered in eleven low-mass X-ray binaries (see van der Klis
\cite{vanderklis97} for a review).  Often, the X-ray power spectra
show twin kHz peaks moving up and down in frequency together.
Sometimes a third kHz peak is detected near a frequency equal to the
separation frequency of the twin peaks, or twice that, suggesting a
beat-frequency interpretation, with the third peak near the neutron
star spin frequency (or twice that).  However, in Sco~X--1 the twin
peak separation varies, which is not consistent with a simple
beat-frequency interpretation (van der Klis et al.
\cite{vanderklis97a}).

In the X-ray burster (Swank et al.  \cite{swank78}; Brandt et al.
\cite{brandt92}) and suspected atoll source (Singh \& Apparao
\cite{singh94}) 4U~0614+09 twin kHz peaks occur (Ford et al.
\cite{ford97}).  The peaks move between 480 and 800~Hz and 520 and
1150~Hz, respectively.  Their separation is consistent with being
constant near 323~Hz.  There has been one 3.3$\sigma$ detection of a
third peak near 328~Hz (in the persistent emission, whereas in the
three other sources where a third peak occured it was seen during
X-ray bursts).

It is yet unclear how the properties of the kHz QPO relate to type and
state of the sources in which they have been observed.  In this paper
we analyze new X-ray timing and spectral data on 4U~0614+09
(preliminary reported previously in van der Klis et al.
\cite{vanderklis96}).  We point out a number of correlations between
the QPO properties and those of the broad-band noise and the X-ray
spectra, and present for the first time an analysis of the photon
energy spectrum of the oscillating flux in kHz QPO.

\section{Observations}

We observed 4U~0614+09 with the proportional counter array (PCA) on
board NASA's Rossi X-Ray Timing Explorer (Bradt, Rothschild, \& Swank
\cite{bradt93}) three times, in February, March, and April 1996
(Table~\ref{tabpow}).  We simultaneously collected 2--60~keV data with
a time resolution of 8~$\mu$s in 8 energy bands, and 16~s in 129
bands.  The background and deadtime corrected count rates were 235,
571, and 256 c/s, respectively.  A change in PCA gain between March
and April affected these values only slightly.  In April, only 3 out
of the 5 PCA detectors were active; the 5 detector count rate was
426~c/s.  We observed no X-ray bursts.

We calculated power spectra of the 8~$\mu$s data and subtracted the
Poisson noise and the Very Large Event window contribution (Zhang et
al.  \cite{zhang95i}; Zhang \cite{zhang95ii}).  We obtained background
measurements from slew and Earth occultations data, and used them to
renormalize all power spectra to fractional rms squared per Hertz (see
van der Klis \cite{vanderklis95}), and to correct the X-ray spectra
obtained from the 16~s data.

\section{Results}

\begin{figure}[h]
\plotfiddle{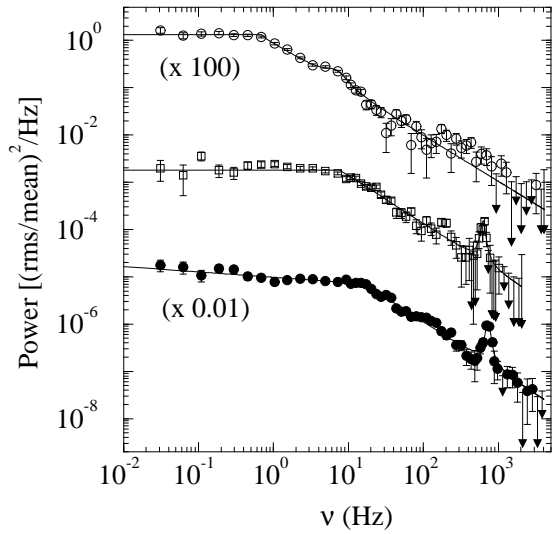}{250pt}{0}{70}{70}{-150}{0}
\caption{The power spectra for the 3 observations
mentioned in the paper. The data from February, March, and April
are represented by open circles, filled circles, and squares
respectively. The significance of the peak at $\sim 200$~Hz in the
power spectrum from April is less than $3 \sigma$ and was not
fitted.  We included a QPO at 6.5~Hz in fitting the February spectrum.
\label{figpow}}
\end{figure}

All power spectra (Fig.~\ref{figpow}) show strong ($\sim$30\% rms)
band-limited noise, which we fitted with a broken power law.  In two
cases we observed kHz QPOs.  These we fitted with Lorentzian peaks.
The fit results are listed in Table~\ref{tabpow}.  The power-law break
frequency varied from 0.7~Hz in February, via 15.4~Hz in March to
6.6~Hz in April.

During March and April strong QPO are present near 727 and 629~Hz,
respectively.  The QPO properties did not vary significantly within
each observation.  We detect no other kHz QPO peaks.  In particular,
any peaks 323~Hz above or below our detected peaks are 3 to 14 times
weaker than these (Table~\ref{tabpow}).  The upper limit on any 328~Hz
peak is 3.5\% (rms; 95\% confidence).

Countrates in March were sufficient to study the photon energy
dependence of the QPO.  We fitted power spectra in 5 energy bands
(1.1--4.3--6.8--8.4--15.5--69.8~keV) with the break frequency, the
power law slopes, and QPO frequency and width fixed to the 2--60~keV
values (none of these varied significantly with photon energy).  As
shown in Fig.~2a the rms amplitude of the QPO rose from 11\% at
3.3~keV via 20\% at 11~keV to (37$\pm$12)\% at 22.8~keV.  Similar
energy dependencies were seen in other sources.  The fractional rms
amplitude of the band-limited noise did not vary significantly as a
function of photon energy.

Fig.~2b shows the energy spectrum of the oscillating flux (QPO rms
amplitude in units of c/s/keV vs.  photon energy).  For reference, we
quote the results of a blackbody fit:  the best fit ($\chi^{2}=14.4$
with 3 dof) has a temperature of (1.56$\pm0.2$~keV) and a radius of
(500$\pm200$~m) (at 3~kpc, Brandt et al. \cite{brandt92}).  The
resolution of the spectrum is low and many spectral models are
consistent with it.  For example, the data can be fitted by a $\sim
2.5$\% temperature variation in a $\sim 1.1$~keV blackbody spectrum
with a radius of 10~km, or ($\chi^{2}=4.9$ with 4 dof) by $\sim 5$\%
optical depth variations in an unsaturated Comptonization spectrum.
(This latter result was obtained by fitting the fractional rms
spectrum.)

To convert count rates to fluxes, we fitted the 2--50~keV energy
spectra with a blackbody plus a power law modified by interstellar
absorption (Table~\ref{tabpow}).  In April and during a dip in soft
color in March (below), a power law alone could fit the spectrum.  The
inferred 2--10~keV fluxes were 0.4, 1.0, and 0.75 $\times 10^{-9}$ erg
cm$^{-2}$ s$^{-1}$ in February, March, and April, respectively; the
2--50~keV fluxes are 50, 30, and 40\% higher, respectively.  As
$N_{\rm H}$ is low absorbed and ``unabsorbed'' fluxes are the same.
We tried various other spectral shapes, with no effect on the derived
fluxes.  Count rate, flux nor power spectra were affected by the soft
color dip.

To compare 4U~0614+09 to confirmed atoll sources (cf.  Hasinger \& van
der Klis \cite{hasinger89}), we produced an X-ray color-color diagram
using the three spectral bands 2--5--10--50~keV (Fig~\ref{color}).  We
corrected for detector gain changes by comparing the count rate ratios
with incident flux ratios obtained in the same bands from the spectral
fits.  Once again we tried different spectral shapes and found
negligible effects on the color corrections.  There was a $\sim 800$~s
dip in soft color starting 5200~s into the March observation.  Apart
from this, the colors within each observation show no significant
changes, with upper limits on any intrinsic color variations of
$\sim$10\%.  Except for the March excursion, this is entirely
consistent with atoll-source island-state behavior.

\section{Discussion}

From color-color diagrams Singh \& Apparao (\cite{singh94}) suggested
4U~0614+09 to be an atoll source.  Our color-color diagrams and
spectral fits, and simultaneous strong band-limited noise with
0.7--15~Hz cutoff frequencies confirm that the correlated X-ray
spectral and timing properties of 4U~0614+09 are those of an atoll
source, in the island state during our observations.  The large
fractional amplitude and low cut-off frequency of the band-limited
noise and the relatively hard X-ray spectra make the source similar to
the atoll sources 4U~1608--52 and 4U~1705--44 in the island state
(Hasinger \& van der Klis \cite{hasinger89}; Langmeier, Hasinger, \&
Tr\"umper \cite{langmeier89}; Yoshida et al.  \cite{yoshida93}; Berger
\& van der Klis \cite{bvk96}), which in turn resemble the black hole
candidates Cyg~X--1 and GX~339--4 in the low and intermediate states
(van der Klis \cite{vanderklis94a}; Berger \& van der Klis
\cite{bvk96}; Belloni et al.  \cite{belloni96}; Crary et al.
\cite{crary96}; M\'endez \& van der Klis \cite{mendez97}).  All these
sources are quite hard.  Their energy spectra fit a soft component
with $kT\sim1$~keV plus a power law with photon index 1.6--2.5.  Their
power spectra show band limited noise with a $\sim0.1-10$~Hz cut-off
frequency that is anticorrelated with the level of the flat top
(Belloni \& Hasinger \cite{belloni90}; M\'endez \& van der Klis
\cite{mendez97}).  This anti-correlation holds also for our power
spectra of 4U~0614+09.  Break frequencies and fractional rms at the
break are fully consistent with the existing relation between these
quantities from other sources (c.f.  Fig.  3 of M\'endez \& van der
Klis \cite{mendez97}).

If as proposed (van der Klis \cite{vanderklis94b}; M\'endez \& van der
Klis \cite{mendez97}), the break frequency of the band limited noise
component here is an indication for $\dot M$, then $\dot M$ increased
from February to March, and then decreased to an intermediate value in
April.  This is consistent with the variations in X-ray flux among our
three observations.  Note, that this is not always the case in similar
sources (e.g., 4U~1608--52; Yoshida et al.  \cite{yoshida93}).

In this picture, we observe no kHz QPO ($<$6\% rms), when $\dot M$ is
lowest, and 11--15\% amplitude QPOs at higher inferred $\dot M$.
Although kHz QPOs often get stronger when inferred $\dot M$ drops (in
4U~1636--53, Wijnands et al.  \cite{wijnandsetal97}; KS~1731--260,
Wijnands \& van der Klis \cite{wijnands97}; 4U~1820--30, Smale, Zhang,
\& White \cite{smale96}), apparently in 4U~0614+09 there is a value of
$\dot M$ below which the QPO becomes weaker.

We now turn to the identification of our QPO peaks.  In our data at
most one peak is present at each time, whereas for similar count rates
Ford et al.  (\cite{ford97}) usually find twin peaks.  The count rate
vs.  QPO frequency relations do not clearly identify our peaks as
either the higher- or the lower-frequency ones (Fig.~\ref{qporate}).
The QPOs amplitudes {\it do} give a clue.  Ford et al.
(\cite{ford97}) found the lower-frequency peak to have an rms
amplitude 0.25--0.75 times that of the higher-frequency peak when the
count rate was near 400~c/s, and 0.5--1.5 that at 600--700~c/s.  If
our peaks are the higher-frequency ones, these ratios are $<$0.3 to
$<0.4$ in our data.  If they are the lower-frequency ones, these
values are $>$2.4 to $>$3.5.  As our countrates are 400--600~c/s, our
peaks are probably the higher-frequency ones.  Then, in our data we
have a positive correlation between QPO frequency and count rate, but
one that does not fit either of the relations observed previously (see
Fig.~\ref{qporate}).

The QPO frequency vs.  count rate diagram of 4U~0614+09
(Fig.~\ref{qporate}) shows that there is no single relation that
describes the correlation between those two quantities.  This means
that either countrate or frequency (or both) does not track $\dot M$
well.  We suggest that count rate is not a good measure of $\dot M$.
Both spectral changes and variations in the anisotropy of the emission
can destroy the expected correspondence between these two quantities.
To the extent that bolometric corrections can be accurately performed,
a conversion to energy flux adjusts for the spectral changes.
However, the experience from the Z sources (at admittedly much higher
$\dot M$) shows that even derived bolometric X-ray fluxes do not
always track $\dot M$ well.  In 4U~1608--52, in a very similar series
of island state observations as the present ones of 4U~0614+09,
Yoshida et al.  (\cite{yoshida93}) found that the bolometric flux did
not exhibit a one-to-one correspondence to cutoff frequency of the
band-limited noise.  Whereas there are many processes that can quite
easily change the observed X-ray count rates, colours, spectral
parameters, and even bolometric fluxes from their original values, the
noise cutoff frequency is not so easily affected.  The noise cutoff
can, in principle, be lowered by scattering delays.  The kHz QPO
frequency, on the other hand, is very likely a direct diagnostic of
the dynamics of the inner flow and is therefore very hard to modify by
any propagation effect.  It is possible, then, that QPO frequency {\it
is} well correlated to $\dot M$, but countrate is not.  It will be of
great interest to see if the correlation suggested by the joint
decrease of the noise cutoff frequency and QPO frequency from March to
April (see Table~\ref{tabpow}) will hold up in future analyses of
other data, or whether a QPO frequency vs.  flux relation will turn
out to be the more reproducible, as this may provide clues to both the
best measure of $\dot M$ in atoll sources in extreme island states
(and perhaps black-hole candidates) and to the physical origin of kHz
QPO and broad-band noise in these systems.

We observed a very strong energy dependence in the kHz QPO but no
energy dependence in the band-limited noise.  One plausible
interpretation of this, namely, that the low energy photons undergo
scattering with a characteristic delay time scale intermediate between
the QPO time scale (1~ms) and that of the noise (0.1--10~s), can
probably be excluded on the basis of the extremely small (20~$\mu$s
and $<$50~$\mu$s) time lags between the kHz QPO signals at different
photon energies recently reported by Vaughan et al.
(\cite{vaughan97}).  (Of course, scattering with shorter
characteristic time scales cannot be excluded and is in fact likely on
other grounds.)  The kHz QPO spectrum which resembles a blackbody
shape with a characteristic temperature of 1.6~keV and radius of
500~m, might indicate an origin associated with a relatively small
area on the neutron star surface, whereas the band-limited noise may
have a {\it different} site of origin, perhaps in the inner disk.
This would be in accordance with the observation that band-limited
noise is a common trait among neutron stars and black holes (van der
Klis \cite{vanderklis94a}), whereas correlated twin kHz QPO peaks are
so far unique to neutron star systems.

\acknowledgements

This work was supported in part by the Netherlands Organization for
Scientific Research (NWO) under grant PGS 78-277 and by the
Netherlands Foundation for research in astronomy (ASTRON) under grant
781-76-017.  MM is a fellow of the Consejo Nacional de Investigaciones
Cient\'{\i}ficas y T\'ecnicas de la Rep\'ublica Argentina.  WHGL
acknowledges support from the National Aeronautics and Space
Administration.  JVP acknowledges support from the National
Aeronautics and Space Administration through contract NAG5-3269.
FKL acknowledges support from NSF, through grant AST 93-15133, and
NASA, through grant 5-2925.

\clearpage

\begin{deluxetable}{lrrrr}
\scriptsize
\tablecaption{Power- and X-ray spectral parameters
\label{tabpow}
}
\tablewidth{0pt}
\tablehead{
\colhead{}          &
\colhead{1996 Feb 26} &
\colhead{1996 Mar 16} &
\colhead{1996 Mar 16} &
\colhead{1996 Apr 13} \nl
&&  (soft color dip)\tablenotemark{a} & (main)\tablenotemark{a}\phn\phn
\phn\phn\phn &\nl
}
\startdata
Start time (UTC)               & 08:53\phm{00000}       & 09:53\phm{00000}                  & 08:26\phm{00000}                  & 09:50\phm{00000}                \nl
Exposure time (ks)             & 5.0\phm{000000}        & 0.8\phm{000000}                   & 4.3\phm{000000}                   & 5.2\phm{000000}                 \nl
\cutinhead{Power spectra}
BLN rms [\%]\tablenotemark{b}\ & $30.2\phn \pm 0.9\phn$ & $29.4\phn \pm \phn1.6\phn$        & $28.1\phn \pm \phn0.7\phn$        & $35.3\phn \pm 1.1\phn$\phn\phn  \nl
$\nu_{\rm break}$ [Hz]         & $ 0.7\phn \pm 0.1\phn$ & $15.4\phn \pm\phn 1.5\phn$        & $15.4\phn \pm \phn0.8\phn$        & $ 6.6\phn \pm 0.5\phn$\phn\phn  \nl
P$_{\rm break}$ rms [\%]\tablenotemark{c}\
                               & $12.2\phn \pm 1.1\phn$ & $ 3.0\phn \pm \phn0.2\phn$            & $ 3.2\phn \pm \phn0.2\phn$        & $ 4.5\phn \pm 0.2\phn$\phn\phn  \nl 
QPO rms [\%]                   & $< 6$\phm{.00000000}   & $11.0\phn \pm \phn1.8\phn$        & $13.9\phn \pm \phn0.7\phn$        & $14.6\phn \pm 0.5\phn$\phn\phn  \nl
Freq. [Hz]                     & \nodata\phm{.00}       & $728\phm{.00} \pm \phn8\phm{.00}$ & $719\phm{.00} \pm \phn5\phm{.00}$ & $629\phm{.00} \pm 4\phm{.0000}$ \nl
FWHM [Hz]                      & \nodata\phm{.00}       & $ 54\phm{.00} \pm    27\phm{.00}$ & $ 99\phm{.00} \pm    14\phm{.00}$ & $98\phm{.00}  \pm 9\phm{.0000}$ \nl
2nd. QPO rms [\%]              & \nodata\phm{.00}       & $< 4$\phm{.000000000}             & $< 4$\phm{.000000000}             & $< 6$\phm{.0000000000}          \nl
$\chi^{2}$/dof\tablenotemark{d}
                               & 284/229\phd\phd        & 270/346\phn\phn                   & 329/346\phn\phn                   & 259/242\phm{..00}               \nl
\cutinhead{Energy Spectra}
N$_{\rm H}$ [10$^{-22}$cm$^{-2}]$\tablenotemark{e}
                               & 0 $(< 1.34)$\phd       & 0 $(< 0.22)$\phd\phn              & 0 $(< 0.10)$\phd\phn              & 0.14 $(< 0.39)$                 \nl
$kT$ [keV]                     & $0.42 \pm 0.05$        & \nodata\phm{.000}                 & $1.10 \pm \phn0.18$               & \nodata\phm{.0000}              \nl
$n$ (photon-index)             & $1.97 \pm 0.03$        & $2.54 \pm \phn0.03$               & $2.41 \pm \phn0.03$               & $2.30 \pm 0.03$\phn\phn         \nl
PL  Flux\tablenotemark{f}      & $0.36 \pm 0.02$        & $1.00 \pm \phn0.05$               & $0.89 \pm \phn0.08$               & $0.75 \pm 0.02$\phn\phn         \nl
BB Flux\tablenotemark{f}       & $0.04 \pm 0.01$        & $<$0.08\phm{0000000}              & $0.12 \pm \phn0.05$               & $<$0.02\phm{00000000}           \nl
$\chi^{2}$/dof\tablenotemark{d}
                               & 103/90\phn\phn         & \phn92/94\phm{000}                & \phn80/92\phm{000}                & \phn95/83\phm{0000}               \nl
\enddata
\tablenotetext{a}{The March observation was divided into 2 parts; see text
for details.}
\tablenotetext{b}{Band limited noise.}
\tablenotetext{c}{Power of the BLN component at $\nu_{\rm break}$.}
\tablenotetext{d}{Number of degrees of freedom.}
\tablenotetext{e}{N$_{\rm H}$ was consistent with zero in all of our fits. The
best-fit values are given first, followed by the upper limits in parentheses.}
\tablenotetext{f}{2--10 keV flux in units of
$10^{-9}$ ergs~cm$^{-2}$~s$^{-1}$. When only an upper limit is given 
for the blackbody flux, this parameter was kept fixed at zero in the 
final fit.}
\tablenotetext{}{Quoted errors represent 90\% confidence intervals for
the fits to the X-ray spectra and 1$\sigma$ confidence intervals for
the fits to the power spectra. Quoted upper limits are 95\%
confidence. A 2\% systematic uncertainty was included in the X-ray
spectral errors to account for the calibration uncertainties (Cui et
al. \cite{cui97}).}
\end{deluxetable}

\clearpage

\clearpage

\begin{figure}[h]
\plotfiddle{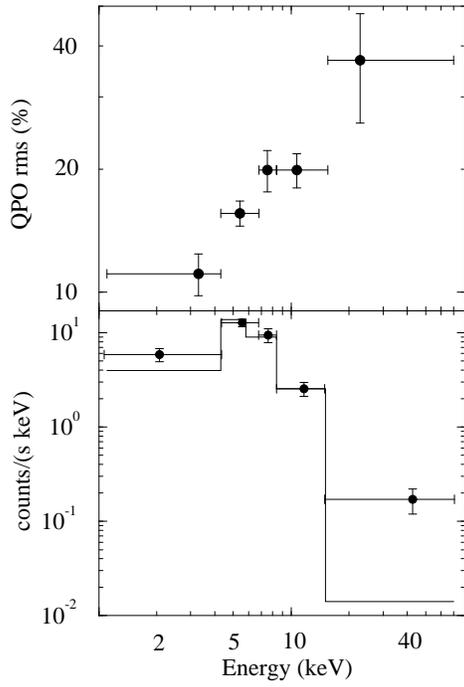}{250pt}{0}{60}{60}{-150}{-40}
\caption{{\it (a)} Fractional rms amplitude vs.  photon
energy spectrum of the QPO from March 1996.  {\it (b)} Energy spectrum
of the oscillating flux, with blackbody fit, for the same data set.
\label{qpoene}}
\end{figure}

\clearpage

\begin{figure}[h]
\plotfiddle{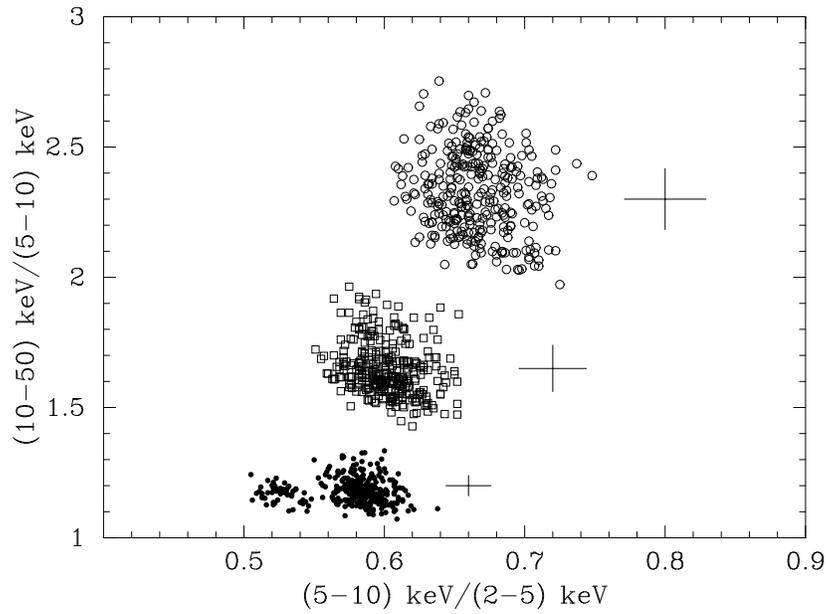}{250pt}{270}{70}{70}{-200}{350}
\caption{Color-color diagram showing all the observations
mentioned in the paper.  Symbols are the same as in Fig.~\ref{figpow}.
Each point represents 16 s of data.  Typical error bars for each
observation are indicated.
\label{color}}
\end{figure}

\clearpage

\begin{figure}[h]
\plotfiddle{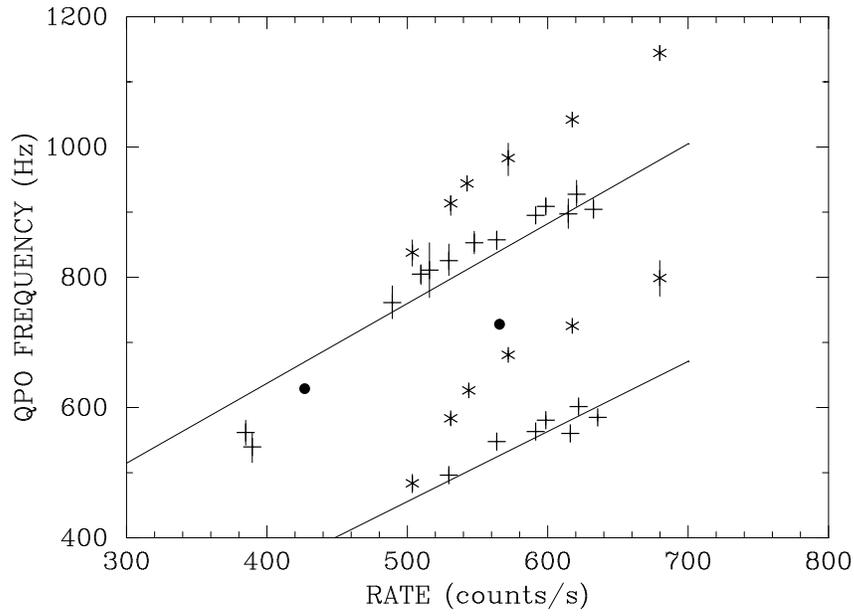}{250pt}{270}{70}{70}{-200}{350}
\caption{The count rate vs.  QPO frequency diagram of
all 4U~0614+09 data reported to date. Pluses, asteriks, and
power-law relations have been taken from Ford et al. (\cite{ford97});
filled circles represent the new data presented in this paper.
\label{qporate}}
\end{figure}

\end{document}